\documentclass{elsarticle}
\usepackage{graphicx}
\usepackage{amssymb}
\usepackage{amsthm}

\newcommand{\beq}{\begin{eqnarray}}
\newcommand{\eeq}{\end{eqnarray}}

\begin{document}

\begin{frontmatter}

\title{Exact sum rules for inhomogeneous drums}
\author{Paolo Amore}
\ead{paolo.amore@gmail.com}
\address{Facultad de Ciencias, CUICBAS, Universidad de Colima,\\
Bernal D\'{i}az del Castillo 340, Colima, Colima, Mexico}

\begin{abstract}
We derive general expressions for the sum rules of the eigenvalues of 
drums of arbitrary shape and arbitrary density, obeying different boundary conditions.
The formulas that we present are a generalization of the analogous formulas for one dimensional
inhomogeneous systems that we have obtained in a previous paper. We also discuss the extension of these
formulas to higher dimensions. 
We show that in the special case of a density depending only on one variable the sum rules 
of any integer order can be expressed in terms of a single series. 
As an application of our result we derive exact sum rules for the a homogeneous circular annulus
with different boundary conditions, for a homogeneous circular sector and for a radially inhomogeneous 
circular annulus with Dirichlet boundary conditions.
\end{abstract}


\end{frontmatter}

\section{Introduction}
\label{sec:intro}

This paper considers the problem of obtaining explicit expressions for the sum rules 
\beq
Z(p) = \sum_{n} \frac{1}{E_n^p}  \ \ , \ \ p=2,3,\dots
\label{zeta}
\eeq
where  $E_n$ are the eigenvalues of the Helmholtz equation 
\beq
(- \Delta) \psi_n(x,y) = E_n \Sigma(x,y) \psi_n(x,y) \ .
\label{helmholtz}
\eeq
over a two dimensional domain $(x,y) \in \Omega$. We assume that $\Omega$ is a domain where 
a orthonormal basis is known (square, circle, etc) since the more general 
problem on an arbitrary domain may be reduced to the form of eq.(\ref{helmholtz}) using a 
conformal map.  Here $n$ is the set of quantum numbers which fully specifies a solution and
$\Sigma(x,y)$ may either be a physical density,  a density obtained from a
conformal map or a composition of the two.

Unfortunately, the evaluation of $Z(p)$ using eq.(\ref{zeta}) requires the knowledge of 
the eigenvalues of the problem, which is granted only in special cases. In a recent paper, 
ref.\cite{Amore12}, we have discussed an alternative approach which uses the representation of
$Z(p)$ directly as the trace of the hermitian operator
\beq
\hat{Q}^p \equiv \left( \sqrt{\Sigma} (-\Delta )^{-1} \sqrt{\Sigma}\right)^p
\eeq
where $p > 1$ is a real exponent. We have proved there that the eigenvalues of $\hat{Q}$ 
are the reciprocals of the eigenvalues of eq.(\ref{helmholtz}) and therefore the trace of 
$\hat{Q}^p$ provides the spectral zeta function associated with the problem of eq.(\ref{helmholtz}).
For integer values of $p$ it is possible to obtain explicit expressions for the trace $\hat{Q}^p$
and therefore for $Z(p)$: we have discussed specific examples in ref.~\cite{Amore12}. 
In the general case of non integer values of $p$, we have shown in ref.\cite{Amore12} that
it is possible to evaluate the trace using perturbation theory; although the expressions obtained 
following this approach are defined for $p > 1$ (in $d$ dimensions, $p>d/2$)
it is possible to perform their analytic continuation to $s<1$: in this way we have 
obtained the Casimir energy of specific systems in one, two and three dimensions. 

There is a large number of works where spectral sum rules for given sytems have been studied: these
include the sum rules for quantum mechanical anharmonic oscillators~\cite{Voros80,Voros99}, 
for potentials $V(r) = g r^p$ ($p>0$, $g>0$) \cite{Steiner85}, for Aharonov-Bohm quantum 
billiards \citep{Berry86,Steiner87}, for the zeroes of Bessel functions~\cite{Steiner85, Elizalde93}, 
for the Selberg's zeta function for compact Riemann surfaces \cite{Steiner87b} (see also \cite{Arendt09}), 
for quantum mechanical one dimensional potentials\cite{Crandall96}, for two dimensional domains 
close to the unit disk \cite{Kvitsinsky96}, for the cardioid and related domains\cite{Dittmar11},
for a $\mathcal{PT}$-symmetric hamiltonian \cite{Mezincescu00,Bender00}.

In this paper we provide a method to calculate the sum rules for the eigenvalues of the negative laplacian in two and higher dimensions, for systems of arbitrary density. The approach that we describe is based on the results of ref.~\cite{Amore12}, which are now expressed in terms of the appropriate Green's functions and used to derive a general formula for the sum rule of arbitrary
order $p > d/2$ ($d$ are the dimensions of the problem). This formula is a generalization
of the analogous formula for one-dimensional inhomogeneous systems obtained in ref.~\cite{Amore13a}, 
where it has been applied to a number of examples  (in the case of a particular inhomogeneous string 
we have been able to obtain sum rules up to order nine). In two dimensions the reference domain 
chosen for our calculation is a rectangle (for $d>2$ we use its generalization), although it can be any domain where a Green's function is available.
In particular in two dimensions, our formula should be compared with the equivalent formula
obtained by Itzykson, Moussa and Luck~\cite{Itzykson86}, who expressed the sum rule 
of a given order $n$  for the Dirichlet eigenvalues of the negative laplacian
on an arbitrary domain $T$ in terms of the Green's function on the upper half plane; 
moreover Dittmar~\cite{Dittmar02} has obtained a formula for the sum rule of order two for the eigenvalues of the negative laplacian on a two dimensional domain both for Dirichlet and Neumann boundary conditions in terms of the Green's functions on a circle (see Theorems 2.1 and 3.3 of ref.~\cite{Dittmar02}). 
The expressions that we derive in this paper apply to sum rules of arbitrary order, to different boundary conditions an both to homogenous and inhomogeneous  systems. 

The paper is organized as follows: in Section \ref{green} we derive the expressions
for the Green's functions on a rectangle of sides $a$ and $b$, with different boundary
conditions; in Section \ref{exact} we obtain a general formula for the sum rules of arbitrary
order in two dimensions; in Section \ref{dim} we discuss the generalization of this formula 
to higher dimensions; in Section \ref{appl} we consider some applications of our formula;
finally, in Section \ref{concl} we draw our conclusions.

\section{Green's functions on a rectangle}
\label{green}

In this section we derive explicit expressions for the Green's functions
of the negative laplacian on a rectangle of sides $a$ and $b$ and obeying
different boundary conditions at the borders. We consider the cases of
Dirichlet, Neumann and periodic bc on the borders of the rectangle, 
of mixed bc, Dirichlet-Neumann, Dirichlet-periodic, Neumann-periodic and
mixed Neumann-Dirichlet boundary condition in one direction and periodic 
in the other direction.

\subsection{Dirichlet boundary conditions}
We consider the rectangular region $(-a/2,a/2) \times (-b/2,b/2)$ and assume 
Dirichlet boundary conditions at the border of this region;  the Green's function 
for the negative laplacian is
\beq
G^{(D)}(x,y;x',y') &=& \sum_{n_x=1}^\infty \sum_{n_y=1}^\infty \frac{\psi^{(D)}_{n_x}(x) \phi^{(D)}_{n_y}(y) 
\psi^{(D)}_{n_x}(x') \phi^{(D)}_{n_y}(y')}{\epsilon^{(D)}_{n_x} + \eta^{(D)}_{n_y} } \ .
\label{greend1}
\eeq

Here $\psi_{n_x}(x)$ and $\phi_{n_y}(y)$ are the Dirichlet eigenfunction of the 1D negative 
laplacian on $x \in (-a/2,a/2)$ and  $y \in (-b/2,b/2)$ respectively:
\beq
\psi_{n_x}(x) \equiv \sqrt{\frac{2}{a}} \ \sin \left( \frac{n_x\pi}{a} (x+a/2)\right) \ \ , \ \ n_x &=& 1,2,\dots \\
\psi_{n_y}(y) \equiv \sqrt{\frac{2}{b}} \ \sin \left( \frac{n_y\pi}{b} (y+b/2)\right) \ \ , \ \ n_y &=& 1,2,\dots
\eeq

$\epsilon^{(D)}_{n_x}$ and  $\eta^{(D)}_{n_y}$ are the Dirichlet eigenvalues of the 1D negative laplacian in the
two directions:
\beq
\epsilon^{(D)}_{n_x} = \frac{n_x^2\pi^2}{a^2} \ \ \  , \ \ \ \eta^{(D)}_{n_y} = \frac{n_y^2\pi^2}{b^2} \ .
\eeq 

Eq.(\ref{greend1}) may be cast in the form:
\beq
G^{(D)}(x,y;x',y') &=& \sum_{n_x=1}^\infty g_{n_x}^{(D)}(y,y') \psi^{(D)}_{n_x}(x)
\psi^{(D)}_{n_x}(x')\ ,
\label{greend2}
\eeq
where (see Eq.(3.169) and problem 2.15 of Ref.~\cite{Jackson}):
\beq
g_{n_x}^{(D)}(y,y') &\equiv& \sum_{n_y=1}^\infty \frac{\phi^{(D)}_{n_y}(y) \phi^{(D)}_{n_y}(y')}{\epsilon^{(D)}_{n_x} + \eta^{(D)}_{n_y} } \nonumber \\
&=& \frac{ \sinh \left( \sqrt{\epsilon_{n_x}^{(D)}} (y_{<}+b/2)\right) 
\sinh \left(\sqrt{\epsilon_{n_x}^{(D)}} (b/2-y_{>})\right) }{\sqrt{\epsilon_{n_x}^{(D)}}  
\sinh \sqrt{\epsilon_{n_x}^{(D)}} b } \ .
\label{eqgd}
\eeq
We have introduced the notation $y_< \equiv \min (y,y')$ and $y_> \equiv \max (y,y')$.

We also define
\beq
G^{(D)}(x,y; x',y') &\equiv& G_+^{(D)}(x,y; x',y') \theta(y-y') \nonumber \\
&+& G_-^{(D)}(x,y; x',y') \theta(y'-y) \ ,
\eeq
where the explicit form of $G_{\pm}^{(D)}(x,y; x',y')$ is easily obtained from Eq.~(\ref{greend2}).

Notice also that
\beq
G_+^{(D)}(x,y; x',y') &=& G_-^{(D)}(x,-y; x',-y') \nonumber \\
G_+^{(D)}(x,y'; x',y) &=& G_-^{(D)}(x,y; x',y') \ .
\nonumber
\eeq

It is also possible to express Eq.(\ref{greend2}) in an alternative form, observing that
\beq
\frac{1}{\sinh \frac{\pi n b}{a} } = 2 \sum_{j=0}^\infty e^{-(2 j+1)\pi n b/a}
\eeq

Upon substitution of this expansion inside Eq.(\ref{greend2}) we have
\beq
G^{(D)}(x,y; x',y') &\equiv& \sum_{j=0}^\infty G_j^{(D)}(x,y; x',y') \nonumber \\
&=&  \frac{1}{4\pi} \sum_{j=0}^\infty \log
\frac{\Omega_j(x_-,x_+,y_-,y_+)}{\Theta_j(x_-,x_+,y_-,y_+)} \nonumber \\
&=&  \frac{1}{4\pi}  \log \prod_{j=0}^\infty\frac{\Omega_j(x_-,x_+,y_-,y_+)}{\Theta_j(x_-,x_+,y_-,y_+)}  \ ,
\label{greend3}
\eeq
where 
\beq
\Omega_j(x_-,x_+,y_-,y_+) &\equiv& 
\left(\cosh \left(\frac{\pi  (2 b j+b-y_{+})}{a}\right)-\cos \left(\frac{\pi x_{-}}{a}\right)\right) \nonumber \\
&\cdot& \left(\cosh \left(\frac{\pi  (2 b j+b+y_{+})}{a}\right)-\cos \left(\frac{\pi  x_{-}}{a}\right)\right) \nonumber \\
&\cdot&   \left(\cosh \left(\frac{\pi  (2 b (j+1)-|y_{-}|)}{a}\right)+\cos \left(\frac{\pi x_{+}}{a}\right)\right) \nonumber \\
&\cdot& \left(\cosh \left(\frac{\pi  (|y_{-}|+2 b j)}{a}\right)+\cos \left(\frac{\pi  x_{+}}{a}\right)\right)
\label{eq2d3} \\
\Theta_j(x_-,x_+,y_-,y_+) &\equiv& \left(\cosh \left(\frac{\pi  (2 b j+b-y_{+})}{a}\right)+\cos \left(\frac{\pi  x_{+}}{a}\right)\right) \nonumber \\
&\cdot& \left(\cosh \left(\frac{\pi  (2 b j+b+y_{+})}{a}\right)+\cos \left(\frac{\pi  x_{+}}{a}\right)\right) \nonumber \\
&\cdot& \left(\cosh\left(\frac{\pi  (2 b (j+1)-|y_{-}|)}{a}\right)-\cos \left(\frac{\pi  x_{-}}{a}\right)\right) \nonumber \\
&\cdot& \left(\cosh \left(\frac{\pi  (|y_{-}|+2 b j)}{a}\right)-\cos \left(\frac{\pi  x_{-}}{a}\right)\right)
\label{eq2d4}
\eeq
and $x_\pm \equiv x_1 \pm x_2$, $y_\pm \equiv y_1 \pm y_2$.

Notice that
\beq
\Theta_0(0,x_+,0,y_+) =  0 \nonumber
\eeq
and therefore $G^{(D)}(x,y;x',y')$ diverges logarithmically when $x_2 \rightarrow x_1$ and $y_2\rightarrow y_1$.

Although the series in Eq.(\ref{greend3}) converges more rapidly than the 
series in Eq.(\ref{greend2}), in general the latter is more appropriate for 
the evaluation of the sum rules.

\subsection{Neumann boundary conditions}

We now come to the calculation of the Green's function for Neumann boundary 
conditions~\footnote{We have not found references reporting the equivalent expressions
of eq.~(\ref{eqgd}) for boundary conditions other than Dirichlet. 
The results that we report for these cases have been derived in the present 
paper following the same approach used for eq.~(\ref{eqgd}).}; 
in this case we have
\beq
G^{(N)}(x,y;x',y') &=& \sum'_{n_x,u_x,n_y,u_y}
\frac{\psi^{(N)}_{n_x,u_x}(x) \phi^{(N)}_{n_y,u_y}(y) \psi^{(N)}_{n_x,u_x}(x') \phi^{(N)}_{n_y,u_y}(y')}{\epsilon^{(N)}_{n_x,u_x} +\eta^{(N)}_{n_y,u_y} } \ ,
\label{greenn1}
\eeq
where 
$\sum'_{n_x,u_x,n_y,u_y}$ is the sum over all possible values of the quantum number,
with the exclusion of the divergent term corresponding to $n_x=n_y=0$ and $u_x=u_y=1$
~\footnote{
$\sum'_{n_x,u_x,n_y,u_y} c_{n_x,u_x,n_y,u_y} \equiv 
\sum_{n_x=1}^\infty \sum_{u_x=1}^2  c_{n_x,u_x,0,1} + \sum_{n_y=1}^\infty \sum_{u_y=1}^2  c_{0,1,n_y,u_y} + 
\sum_{n_x=1}^\infty \sum_{u_x=1}^2 \sum_{n_y=1}^\infty \sum_{u_y=1}^2 c_{n_x,u_x,n_y,u_y}$.}

Here $\epsilon_{n_x,u_x}^{(N)}$ and $\eta_{n_y,u_y}^{(N)}$ are the Neumann eigenvalues of the 
negative laplacian in the two orthogonal directions;
\beq
\epsilon_{n_x,u_x}^{(N)} &=& \left\{ \begin{array}{ccccc}
\frac{4 n_x^2\pi^2}{a^2}     & , & u_x=1 & , & n_x = 0,1,2,\dots \\
\frac{(2 n_x-1)^2\pi^2}{a^2} & , & u_x=2 & , & n_x = 1,2,\dots\\
\end{array}
\right.   \\
\eta_{n_y,u_y}^{(N)} &=& \left\{ \begin{array}{ccccc}
\frac{4 n_y^2\pi^2}{b^2}     & , &  u_y=1 & , & n_y = 0,1,2,\dots \\
\frac{(2 n_y-1)^2\pi^2}{b^2} & , &  u_y=2 & , & n_y = 1,2,\dots \\
\end{array}
\right. \nonumber
\eeq
and $\psi_{n_x,u_x}^{(N)}(x)$ and $\phi_{n_y,u_y}^{(N)}(y)$ are its eigenfunctions:
\beq
\psi_{n_x,u_x}^{(N)}(x) &=& \left\{\begin{array}{ccc}
\sqrt{\frac{1}{a}} & , & n_x = 0 \ , \ u_x=1 \\
\sqrt{\frac{2}{a}} \cos \frac{2 n_x \pi x}{a} & , & n_x>0 \ , \ u_x=1 \\
\sqrt{\frac{2}{a}} \sin \frac{(2 n_x-1) \pi x}{a} & , & n_x \geq 0 \ , \ u_x=2 \\
\end{array}
\right. \\
\phi_{n_y,u_y}^{(N)}(y) &=& \left\{\begin{array}{ccc}
\sqrt{\frac{1}{b}} & , & n_y = 0 \ , \ u_y=1 \\
\sqrt{\frac{2}{b}} \cos \frac{2 n_y \pi y}{b} & , & n_y>0 \ , \ u_y=1 \\
\sqrt{\frac{2}{b}} \sin \frac{(2 n_y-1) \pi y}{b} & , & n_y \geq 0 \ , \ u_y=2 \\
\end{array}
\right. 
\eeq

In this case Eq.~(\ref{greenn1}) may be expressed as
\beq
G^{(N)}(x,y;x',y') &=& \frac{1}{a} g_{0,1}^{(N)}(y,y') \nonumber \\
&+& \sum_{n_x=1}^\infty \sum_{u_x=1}^2 g_{n_x,u_x}^{(N)}(y,y') 
\psi^{(N)}_{n_x,u_x}(x) \psi^{(N)}_{n_x,u_x}(x')
\label{greenn2}
\eeq
where $g_{n_x,u_x}^{(N)}(y,y')$ is the analogous of Eq.(\ref{eqgd}) for Neumann boundary 
conditions:
\beq
g_{n_x,u_x}^{(N)}(y,y') &\equiv& 
\frac{1-\delta_{n_x,0}}{b \epsilon^{(N)}_{n_x,u_x} } + 
\sum_{u_y=1}^2 \sum_{n_y=1}^\infty \frac{\phi^{(N)}_{n_y,u_y}(y) \phi^{(N)}_{n_y,u_y}(y')}{\epsilon^{(N)}_{n_x,u_x} + \eta^{(N)}_{n_y,u_y} } \nonumber .
\eeq

For $n_x=0$ and $u_x=1$, $g_{n_x,u_x}^{(N)}(y,y')$ reduces to the one-dimensional Green's function with Neumann boundary conditions, which is reported in Ref.~\cite{Amore13a}:
\beq
g_{0,1}^{(N)}(y,y') &=& \frac{b^2+6 b (y_<-y_>)+6 \left(y_>^2+y_<^2\right)}{12 b} 
\label{eqgn1} \ .
\eeq
For $n_x>0$ and $u_x=1,2$ we have
\beq
g_{n_x,u_x}^{(N)}(y,y') &=& \frac{
\cosh \left(\sqrt{\eta^{(N)}_{n_x,u_x}}  \frac{(b-2 y_>)}{2}\right) 
\cosh \left(\sqrt{\eta^{(N)}_{n_x,u_x}}  \frac{(b+2 y_<)}{2}\right)}{\sqrt{\eta^{(N)}_{n_x,u_x}}  
\sinh \left(\sqrt{\eta^{(N)}_{n_x,u_x}} b \right) }  \ .
\label{eqgn2}
\eeq

\subsection{Dirichlet-Neumann boundary conditions}

We now come to the calculation of the Green's function for mixed Dirichlet-Neumann boundary conditions; 
in this case we have
\beq
G^{(DN)}(x,y;x',y') &=& \sum_{n_x,n_y,u_y}
\frac{\psi^{(D)}_{n_x}(x) \phi^{(N)}_{n_y,u_y}(y) \psi^{(D)}_{n_x}(x') \phi^{(N)}_{n_y,u_y}(y')}{\epsilon^{(D)}_{n_x} +\eta^{(N)}_{n_y,u_y} } \ ,
\label{greedn1}
\eeq
where 
\beq
\sum_{n_x,n_y,u_y} c_{n_x,n_y,u_y} = \sum_{n_x=1}^\infty c_{n_x,0,1} + 
\sum_{n_x=1}^\infty  \sum_{n_y=1}^\infty \sum_{u_y=1}^2 c_{n_x,n_y,u_y} \label{eqsum} \ .
\eeq

We may express eq.(\ref{greedn1}) as:
\beq
G^{(DN)}(x,y;x',y') &=& \frac{g_{0,1}^{(DN)}(x,x')}{b} + \sum_{n_y=1}^\infty \sum_{u_y=1}^2 g_{n_y,u_y}^{(DN)}(x,x')
\phi^{(N)}_{n_y,u_y}(y) \phi^{(N)}_{n_y,u_y}(y')
\label{greedn1a}
\eeq
where
\beq
g_{n_y,u_y}^{(DN)}(x,x') &\equiv& \sum_{n_x=1}^\infty 
\frac{\psi^{(D)}_{n_x}(x) \psi^{(D)}_{n_x}(x')}{\epsilon^{(D)}_{n_x} + \eta^{(N)}_{n_y,u_y} } 
\eeq

For $n_y=0$ and $u_y=1$ $g_{n_y,u_y}^{(DN)}(x,x')$ is the 1D Green's function
for Dirichlet bc reported in Ref.\cite{Amore13a}:
\beq
g_{0,1}^{(DN)}(x,x') &=& \frac{(a-2 x_>) (a+2 x_<)}{4a}
\eeq
while for $n_y>0$ and $u_y=1,2$
\beq
g_{n_y,u_y}^{(DN)}(x,x') &=& \frac{
\sinh \left(\sqrt{\eta^{(N)}_{n_y,u_y}} (x_> - \frac{a}{2})\right) \sinh
   \left(\sqrt{\eta^{(N)}_{n_y,u_y}} (\frac{a}{2}+ x_<)\right)}{\sqrt{\eta^{(N)}_{n_y,u_y}} \ 
   \sinh\left(\sqrt{\eta^{(N)}_{n_y,u_y}} \ a \right) } 
\eeq
for $n_y>0$ and $u_y=1,2$.

Alternatively we may write the Green's function as
\beq
G^{(DN)}(x,y;x',y') &=& \sum_{n_x=1}^\infty \tilde{g}_{n_x}^{(DN)}(y,y')
\psi^{(D)}_{n_x}(x)\psi^{(D)}_{n_x}(x') 
\label{greedn1b}
\eeq
where
\beq
\tilde{g}_{n_x}^{(DN)}(y,y') &\equiv&  \frac{1}{b \ \epsilon^{(D)}_{n_x}} + \sum_{u_y=1}^2 \sum_{n_y=1}^\infty 
\frac{\phi^{(N)}_{n_y,u_y}(y) \phi^{(N)}_{n_y,u_y}(y')}{\epsilon^{(D)}_{n_x} + \eta^{(N)}_{n_y,u_y} } 
\nonumber \\
&=& \frac{ \cosh \left( \sqrt{\epsilon_{n_x}^{(D)}} (b/2-y_>)\right) 
\cosh \left(\sqrt{\epsilon_{n_x}^{(D)}} (b/2+ y_<)\right)}{\sqrt{\epsilon_{n_x}^{(D)}} \ \sinh(b \sqrt{\epsilon_{n_x}^{(D)}})}
\eeq

\subsection{Periodic boundary conditions}

We now come to the calculation of the Green's function for periodic boundary conditions; 
in this case we have
\beq
G^{(P)}(x,y;x',y') &=& \sum_{n_x,u_x,n_y,u_y}^\prime
\frac{\psi^{(P)}_{n_x,u_x}(x) \phi^{(P)}_{n_y,u_y}(y) \psi^{(P)}_{n_x,u_x}(x') \phi^{(P)}_{n_y,u_y}(y')}{\epsilon^{(P)}_{n_x,u_x} +\eta^{(P)}_{n_y,u_y} } \ ,
\label{greenp1}
\eeq
where we have excluded from the sum the divergent term corresponding to $n_x=n_y=0$ and $u_x=u_y=1$ 
as for the case of Neumann bc.

Here $\epsilon_{n_x,u_x}^{(P)}$ and $\eta_{n_y,u_y}^{(P)}$ are the Neumann eigenvalues of the 
negative laplacian in the two orthogonal directions;
\beq
\epsilon_{n_x,u_x}^{(P)} = \frac{4 \pi^2 n_x^2}{a^2}   \ \ \ , \ \ \
\eta_{n_y,u_y}^{(N)} = \frac{4 \pi^2 n_y^2}{b^2} 
\eeq
and $\psi_{n_x,u_x}^{(P)}(x)$ and $\phi_{n_y,u_y}^{(P)}(y)$ are its eigenfunctions:
\beq
\psi_{n_x,u_x}^{(P)}(x) &=& \left\{\begin{array}{ccc}
\sqrt{\frac{1}{a}} & , & n_x = 0 \ , \ u_x=1 \\
\sqrt{\frac{2}{a}} \cos \frac{2 n_x \pi x}{a} & , & n_x>0 \ , \ u_x=1 \\
\sqrt{\frac{2}{a}} \sin \frac{2 n_x \pi x}{a} & , & n_x \geq 0 \ , \ u_x=2 \\
\end{array}
\right. \\
\phi_{n_y,u_y}^{(P)}(y) &=& \left\{\begin{array}{ccc}
\sqrt{\frac{1}{b}} & , & n_y = 0 \ , \ u_y=1 \\
\sqrt{\frac{2}{b}} \cos \frac{2 n_y \pi y}{b} & , & n_y>0 \ , \ u_y=1 \\
\sqrt{\frac{2}{b}} \sin \frac{2 n_y \pi y}{b} & , & n_y \geq 0 \ , \ u_y=2 \\
\end{array}
\right. 
\eeq

In this case Eq.~(\ref{greenp1}) may be expressed as
\beq
G^{(P)}(x,y;x',y') &=&  \frac{g_{0,1}^{(P)}(y,y')}{a}
+ \sum_{n_x=1}^\infty \sum_{u_x=1}^2 g_{n_x,u_x}^{(P)}(y,y') 
\psi^{(P)}_{n_x,u_x}(x) \psi^{(P)}_{n_x,u_x}(x')
\label{greenp2}
\eeq
where 
\beq
g_{n_x,u_x}^{(P)}(y,y') &\equiv& \frac{1-\delta_{n_x,0}}{b \ \epsilon^{(P)}_{n_x,u_x} } + 
\sum_{u_y=1}^2 \sum_{n_y=1}^\infty \frac{\phi^{(P)}_{n_y,u_y}(y) 
\phi^{(P)}_{n_y,u_y}(y')}{\epsilon^{(P)}_{n_x,u_x} + \eta^{(P)}_{n_y,u_y} } \nonumber .
\eeq

For $n_x=0$ and $u_x=1$, $g_{n_x,u_x}^{(P)}(y,y')$ reduces to the one-dimensional Green's function with periodic boundary conditions reported in ref.~\cite{Amore13a}:
\beq
g_{0,1}^{(P)}(y,y') &=& \frac{b^2 - 6 b |y-y'|+6 \left(y-y'\right)^2}{12 b} 
\label{eqgp1} \ .
\eeq

For $n_x>0$ and $u_x=1,2$ we have
\beq
g_{n_x,u_x}^{(P)}(y,y') &=& 
\frac{\cosh \left(\sqrt{\epsilon_{n_x,u_x}^{(P)}} (|y-y'| -b/2)
\right)}{2 \sqrt{\epsilon_{n_x,u_x}^{(P)}} \sinh\left(\frac{b \sqrt{\epsilon_{n_x,u_x}^{(P)}}}{2}\right) }
\eeq

\subsection{Dirichlet-Periodic boundary conditions}

We now come to the calculation of the Green's function for mixed Dirichlet-periodic boundary conditions; in this case we have
\beq
G^{(DP)}(x,y;x',y') &=& \sum_{n_x,n_y,u_y} 
\frac{\psi^{(D)}_{n_x}(x) \phi^{(P)}_{n_y,u_y}(y) \psi^{(D)}_{n_x}(x') \phi^{(P)}_{n_y,u_y}(y')}{\epsilon^{(D)}_{n_x} +\eta^{(P)}_{n_y,u_y} } \ .
\label{greedp1}
\eeq
where $\sum_{n_x,n_y,u_y}$ is defined in eq.(\ref{eqsum}).

We may express this Green's functions as
\beq
G^{(DP)}(x,y;x',y') &=& \frac{g_{0,1}^{(DP)}(x,x')}{b} \nonumber \\
&+& \sum_{n_y=1}^\infty \sum_{u_y=1}^2 g_{n_y,u_y}^{(DP)}(x,x')
\phi^{(P)}_{n_y,u_y}(y) \phi^{(P)}_{n_y,u_y}(y')
\label{greedp1a}
\eeq
where
\beq
g_{n_y,u_y}^{(DP)}(x,x') &\equiv& 
\sum_{n_x=1}^\infty \frac{\psi^{(D)}_{n_x}(x) \psi^{(D)}_{n_x}(x')}{\epsilon^{(D)}_{n_x} + \eta^{(P)}_{n_y,u_y} } 
\eeq

For $n_y=0$ and $u_y=1$, $g_{n_y,u_y}^{(DP)}(x,x')$ reduces to the 1D Green's function
for Dirichlet bc:
\beq
g_{0,1}^{(DP)}(x,x') &=& \frac{(a-2 x_>) (a+2 x_<)}{4a}
\eeq

For $n_y>0$ and $u_y=1,2$
\beq
g_{n_y,u_y}^{(DP)}(x,x') &=& 
\frac{\sinh \left(\sqrt{\eta^{(P)}_{n_y,u_y}} (x_> - \frac{a}{2})\right) \sinh
   \left(\sqrt{\eta^{(P)}_{n_y,u_y}} (\frac{a}{2}+ x_<)\right)}{\sqrt{\eta^{(P)}_{n_y,u_y}} \ 
   \sinh\left(\sqrt{\eta^{(P)}_{n_y,u_y}} \ a \right) } 
\eeq

Alternatively we may write the Green's function as
\beq
G^{(DP)}(x,y;x',y') &=& \sum_{n_x=1}^\infty \tilde{g}_{n_x}^{(DP)}(y,y')
\psi^{(D)}_{n_x}(x)\psi^{(D)}_{n_x}(x') 
\label{greedp1b}
\eeq
where
\beq
\tilde{g}_{n_x}^{(DP)}(y,y') &\equiv& \frac{1}{b \ \epsilon^{(D)}_{n_x}} + \sum_{u_y=1}^2 \sum_{n_y=1}^\infty \frac{\phi^{(P)}_{n_y,u_y}(y) \phi^{(P)}_{n_y,u_y}(y')}{\epsilon^{(D)}_{n_x} + \eta^{(P)}_{n_y,u_y} } \nonumber \\
 &=& \frac{\cosh \left(\sqrt{\epsilon_{n_x}^{(D)}} (|y-y'| -b/2)
\right)}{2 \sqrt{\epsilon_{n_x}^{(D)}} \sinh\left(\frac{b \sqrt{\epsilon_{n_x}^{(D)}}}{2}\right) }
\eeq

\subsection{Neumann-Periodic boundary conditions}

We now come to the calculation of the Green's function for mixed Neumann-periodic boundary conditions:
\beq
G^{(NP)}(x,y;x',y') &=& \sum_{n_x,u_x,n_y,u_y}^\prime
\frac{\psi^{(N)}_{n_x,u_x}(x) \phi^{(P)}_{n_y,u_y}(y) \psi^{(N)}_{n_x,u_x}(x') \phi^{(P)}_{n_y,u_y}(y')}{\epsilon^{(N)}_{n_x,u_x} +\eta^{(P)}_{n_y,u_y} } \ .
\label{greennp1}
\eeq
where $\sum_{n_x,u_x,n_y,u_y}^\prime$ has been defined earlier for the case of Neumann boundary conditions.

We may express the Green's functions as
\beq
G^{(NP)}(x,y;x',y') &=& \frac{g_{01}^{(NP)}(x,x')}{b} + \sum_{n_y=1}^\infty \sum_{u_y=1}^2 g_{n_y,u_y}^{(NP)}(x,x')
\phi^{(P)}_{n_y,u_y}(y) \phi^{(P)}_{n_y,u_y}(y')
\label{greennp1a}
\eeq
where
\beq
g_{n_y,u_y}^{(NP)}(x,x') &\equiv&  \frac{1 - \delta_{n_y,0}}{a \ \eta^{(P)}_{n_y,u_y} } + 
\sum_{n_x=1}^\infty \sum_{u_x=1}^2 \frac{\psi^{(N)}_{n_x,u_x}(x) \psi^{(N)}_{n_x,u_x}(x')}{\epsilon^{(N)}_{n_x,u_x} + \eta^{(P)}_{n_y,u_y} }  \ .
\eeq

For $n_y=0$ and $u_y=1$, $g_{n_y,u_y}^{(NP)}(x,x')$ reduces to the 1D Green's function
for Neumann bc reported in Ref.~\cite{Amore13a} 
\beq
g_{0,1}^{(NP)}(x,x') &=& \frac{a^2-6a |x-x'|+6 (x^2+{x'}^2)}{12a} \ .
\eeq

For $n_y>0$ and $u_y=1,2$
\beq
g_{n_y,u_y}^{(NP)}(x,x') &=&  \frac{
\cosh \left(\sqrt{\eta^{(P)}_{n_y,u_y}}  \frac{(a-2 x_>)}{2}\right) 
\cosh \left(\sqrt{\eta^{(P)}_{n_y,u_y}}  \frac{(a+2 x_<)}{2}\right)}{\sqrt{\eta^{(P)}_{n_y,u_y}}  
\sinh \left(\sqrt{\eta^{(P)}_{n_y,u_y}} a \right) }  \ .
\eeq

Alternatively we may write the Green's function as
\beq
G^{(NP)}(x,y;x',y') &=& \frac{\tilde{g}_{01}^{(NP)}(y,y')}{a}
 + \sum_{n_x=1}^\infty \sum_{u_x=1}^2 \tilde{g}_{n_x,u_x}^{(NP)}(y,y')
\psi^{(N)}_{n_x,u_x}(x)\psi^{(N)}_{n_x,u_x}(x') 
\label{greennp1b}
\eeq
where
\beq
\tilde{g}_{n_x,u_x}^{(NP)}(y,y') &\equiv&  \frac{1-\delta_{n_x,0}}{b \ \epsilon^{(N)}_{n_x,u_x}} + \sum_{u_y=1}^2 \sum_{n_y=0}^\infty \frac{\phi^{(P)}_{n_y,u_y}(y) \phi^{(P)}_{n_y,u_y}(y')}{\epsilon^{(N)}_{n_x,u_x} + \eta^{(P)}_{n_y,u_y} }  \ .
\eeq

For $n_y=0$ and $u_y=1$, $\tilde{g}_{n_y,u_y}^{(NP)}(x,x')$ reduces to the 1D Green's function
for periodic bc reported in Ref.~\cite{Amore13a} 
\beq
g_{0,1}^{(NP)}(y,y') &=& \frac{b^2 - 6 b |y-y'|+6 \left(y-y'\right)^2}{12 b}  \ .
\eeq

For $n_y>0$ and $u_y=1,2$
\beq
\tilde{g}_{n_x,u_x}^{(NP)}(y,y') &=& 
\frac{\cosh \left(\sqrt{\epsilon_{n_x,u_x}^{(N)}} (|y-y'| -b/2)
\right)}{2 \sqrt{\epsilon_{n_x,u_x}^{(N)}} \sinh\left(\frac{b \sqrt{\epsilon_{n_x,u_x}^{(N)}}}{2}\right) } \nonumber \ .
\eeq

\subsection{Neumann-Dirichlet-Periodic boundary conditions}

We now come to the calculation of the Green's function for mixed Neumann-Dirichlet-periodic boundary conditions:
in this case we assume Neumann bc at $x=-a/2$, Dirichlet bc at $x=+a/2$ and periodic boundary conditions
at $y = \pm b/2$.

We have
\beq
G^{(NDP)}(x,y;x',y') &=& \sum_{n_x,n_y,u_y} 
\frac{\psi^{(ND)}_{n_x}(x) \phi^{(P)}_{n_y,u_y}(y) \psi^{(ND)}_{n_x}(x') \phi^{(P)}_{n_y,u_y}(y')}{\epsilon^{(ND)}_{n_x} +\eta^{(P)}_{n_y,u_y} } 
\label{green_ndp1}
\eeq
where $\sum_{n_x,n_y,u_y}$ is defined in eq.(\ref{eqsum}).

We have
\beq
\psi_{n_x}^{(ND)}(x) = \sqrt{\frac{2}{a}} \sin \left(\frac{\pi  (2 n_x-1) (3 a+2 x)}{4 a}\right)
\eeq
and
\beq
\epsilon^{(ND)}_{n_x} = \frac{(2 n_x-1)^2 \pi^2}{4a^2}
\eeq

We may cast the Green's function of Eq.(\ref{green_ndp1}) as
\beq
G^{(NDP)}(x,y;x',y') &=& \frac{g_{0,1}^{(NDP)}(x,x')}{b} + \sum_{n_y=1}^\infty \sum_{u_y=1}^2 g_{n_y,u_y}^{(NDP)}(x,x') \phi^{(P)}_{n_y,u_y}(y)) \phi^{(P)}_{n_y,u_y}(y')
\label{green_ndp2}
\eeq
where
\beq
g_{n_y,u_y}^{(NDP)}(x,x') &\equiv&  \sum_{n_x=1}^\infty \frac{\psi^{(ND)}_{n_x}(x) \psi^{(ND)}_{n_x}(x')}{\epsilon^{(ND)}_{n_x} + \eta^{(P)}_{n_y,u_y} } \ .
\eeq

For $n_y=0$ and $u_y=1$ we have the one-dimensional Green's function for mixed Neumann-Dirichlet boundary conditions:
\beq
g_{0,1}^{(NDP)}(x,x') &=& (- x_>+a/2) 
\eeq

For $n_y>0$ and $u_y=1,2$ we have
\beq
g_{n_y,u_y}^{(NDP)}(x,x') &=& \frac{{\rm sech}\left(a \sqrt{\epsilon_{n_y,u_y}^{(P)}}\right) \sinh \left(\frac{1}{2}
   \sqrt{\epsilon_{n_y,u_y}^{(P)}} (a-2 x_>)\right) \cosh \left(\frac{1}{2}
   \sqrt{\epsilon_{n_y,u_y}^{(P)}} (a+2 x_<)\right)}{\sqrt{\epsilon_{n_y,u_y}^{(P)}}}
\eeq

Alternatively we may cast the Green's function of Eq.(\ref{green_ndp1}) as
\beq
G^{(NDP)}(x,y;x',y') &=& \sum_{n_x=1}^\infty  \tilde{g}_{n_x}^{(NDP)}(y,y') \psi^{(ND)}_{n_x}(x)) \psi^{(ND)}_{n_x}(x')
\label{green_ndp3}
\eeq
where
\beq
\tilde{g}_{n_x}^{(NDP)}(y,y') &\equiv&  \frac{1}{b \ \epsilon^{(ND)}_{n_x} }  + 
\sum_{n_y=1}^\infty \sum_{u_y=1}^2 \frac{\phi^{(P)}_{n_y u_y}(y) \phi^{(P)}_{n_y u_y}(y')}{\epsilon^{(ND)}_{n_x} + \eta^{(P)}_{n_y,u_y} } \nonumber \\
&=& \frac{\cosh \left(\sqrt{\epsilon_{n_x}^{(ND)}} (|y-y'| -b/2)
\right)}{2 \sqrt{\epsilon_{n_x}^{(ND)}} \sinh\left(\frac{b \sqrt{\epsilon_{n_x}^{(ND)}}}{2}\right) }
\eeq

\subsection{Dirichlet-Neumann-Periodic boundary conditions}

In this case we assume Dirichlet bc at $x=-a/2$, Neumann bc at $x=+a/2$ and periodic boundary conditions
at $y = \pm b/2$. This case is trivial since the basis with 
\beq
\psi_{n_x}^{(DN)}(x) = \psi_{n_x}^{(ND)}(-x)
\eeq

Therefore
\beq
g_{n_y,u_y}^{(DNP)}(x,x') &=& g_{n_y,u_y}^{(NDP)}(-x,-x') 
\eeq

\section{Exact sum rules}
\label{exact}

In Ref.~\cite{Amore13a} we have derived a set of rules which allow one to obtain explicit expressions for the sum rules for the eigenvalues of an inhomogeneous string:
\beq
Z_n = \sum_{p} \frac{1}{E_p^n} \  ,
\eeq
with $n=1,2,\dots$.

These rules may be easily generalized to higher dimensions: in particular
in two dimensions the sum rule of order $n$ may be expressed as
\beq
Z^{(D)}_n &=& \int d^2R_1 \int d^2R_2 \dots \int d^2R_n \ G({\bf R}_n; {\bf R}_1)  \nonumber \\
&\cdot&
\prod_{i=1}^{n-1} \left[ G({\bf R}_i; {\bf R}_{i+1})\right]
\prod_{i=1}^n \left[ \Sigma({\bf R}_i) \right]
\eeq
where ${\bf R}_i \equiv (x_i,y_i)$ and $|x_i| \leq a/2$, $|y_i| \leq b/2$. 
Here $G({\bf R}_i; {\bf R}_j)$ is the 2D Green's function obeying the appropriate
bc and $\Sigma({\bf R}_i)$ is a density~\footnote{Notice that $G({\bf R}, {\bf R}')$ is
the Green's function on any domain where a basis is known, such as a rectangle or a circle.}.

As done in Ref.~\cite{Amore13a} it is convenient to cast this expression in an alternative form, using the "y-ordered" Green's function:
\beq
Z^{(D)}_n &=& \int_{-a/2}^{a/2} dx_1 \int_{-a/2}^{a/2} dx_2 \dots \int_{-a/2}^{a/2} dx_n
\int_{-b/2}^{b/2} dy_1 \ \int_{-b/2}^{y_1} dy_2 \dots \int_{-b/2}^{y_{n-1}} dy_n \nonumber \\
&\cdot& \mathcal{G}\left({\bf R}_1,{\bf R}_2, \dots, {\bf R}_n\right) \prod_{i=1}^n \left[ \Sigma({\bf R}_i) \right] \ ,
\eeq
where
\beq
\mathcal{G}\left({\bf R}_1,{\bf R}_2, \dots, {\bf R}_n\right)  \equiv  \left\{
G({\bf R}_n; {\bf R}_1) \prod_{i=1}^{n-1} \left[ G({\bf R}_i; {\bf R}_{i+1})
\right] \right\}_{\mathcal{P}}
\eeq
and
\beq
\left\{ f({\bf R}_1, \dots, {\bf R}_n) \right\}_{\mathcal{P}} 
\equiv \sum_{\rm permutations} f({\bf R}_{p_1}, \dots, {\bf R}_{p_n})  \ .
\eeq

For example, the expressions for $\mathcal{G}$ up to order $4$ are:
\beq
\mathcal{G}\left({\bf R}_1\right) &=& G_+(x_1,y_1; x_1,y_1) \nonumber \\
\mathcal{G}\left({\bf R}_1,{\bf R}_2\right) &=& 2 \ \left[G_+(x_1,y_1; x_2,y_2)\right]^2 
\nonumber \\
\mathcal{G}\left({\bf R}_1,{\bf R}_2,{\bf R}_3\right) &=& 6 \ G_+(x_1,y_1; x_2,y_2) 
G_+(x_2,y_2; x_3,y_3) G_+(x_1,y_1; x_3,y_3) \nonumber \\
\mathcal{G}\left({\bf R}_1,{\bf R}_2,{\bf R}_3,{\bf R}_4\right) &=& 
8 \left[ G_+(x_1,y_1; x_2,y_2) G_+(x_1,y_1; x_4,y_4) G_+(x_2,y_2; x_3,y_3) G_+(x_3,y_3; x_4,y_4) \right. \nonumber  \\
&+& \left. G_+(x_1,y_1; x_3,y_3) G_+(x_1,y_1; x_4,y_4) G_+(x_2,y_2; x_3,y_3) G_+(x_2,y_2; x_4,y_4) \right. \nonumber \\
&+& \left. G_+(x_1,y_1; x_2,y_2) G_+(x_1,y_1; x_3,y_3) G_+(x_2,y_2; x_4,y_4) G_+(x_3,y_3; x_4,y_4) \right] \nonumber
\eeq

We can therefore calculate $Z(n)$ with the diagrammatic rules:
\begin{itemize}
\item Draw n points ${\bf R}_1, \dots, {\bf R}_n$ on a line; 
\item Connect each point to any two other points in all possible inequivalent ways excluding
the disconnected diagrams and the diagrams corresponding to a cyclic permutation of the points;  
\item Associate a density $\Sigma({\bf R}_i)$ at each point ${\bf R}_i$ ($i=1,\dots, n$);
\item Associate a factor $G_+({\bf R}_i; {\bf R}_j)$ to each line connecting ${\bf R}_i$ to ${\bf R}_j$ ($i<j$);
\item Multiply the result by a factor $2n$, corresponding to the $n$ cyclic permutations
of each inequivalent configuration and to the 2 possible directions in which each diagram can
be traveled;
\item Integrate the expression obtained from the steps above over the internal
points: 
\beq
\int_{-a/2}^{a/2} dx_1 \int_{-a/2}^{a/2} dx_2 \dots \int_{-a/2}^{a/2} dx_n
\int_{-b/2}^{b/2} dy_1 \ \int_{-b/2}^{y_1} dy_2 \dots \int_{-b/2}^{y_{n-1}} dy_n \nonumber
\eeq
\end{itemize}

It is easy to convince oneself that for the sum rule of order $n$ there are $n!/2 n = (n-1)!/2$ 
independent diagrams (for $n>2$). 

These rules imply that the sum rule of order $n$ for a general density contains $n$ series,
one for each factor $G_+({\bf R}_i; {\bf R}_j)$ appearing in the expression; however,
in the case of a density which depends only on one variable the orthogonality of the eigenfunctions
along the homogeneous direction allow to reduce the multiple series to a single series, leaving
only the integrals along the direction where the density varies.

Assuming for simplicity that $\Sigma= \Sigma(y)$, and expressing for the Green's function as
\beq
G_+(x,y;x',y') &=& \sum_{p}^\infty g_{p}^{(+)}(y,y')\psi_p(x) \psi_{p}(x')
\eeq
we obtain:
\beq
Z^{(D)}_n &=& 
\sum_p \int_{-b/2}^{b/2} dy_1 \ \int_{-b/2}^{y_1} dy_2 \dots \int_{-b/2}^{y_{n-1}} dy_n \ 
\mathcal{Q}\left(y_1,y_2, \dots, y_n\right) \prod_{i=1}^n \left[ \Sigma(y_i) \right] \ ,
\eeq
where
\beq
\mathcal{Q}\left(y_1,y_2, \dots, y_n\right)  \equiv  \left\{
g_p(y_n; y_1) \prod_{i=1}^{n-1} \left[ g_p(y_i; y_{i+1})\right] \right\}_{\mathcal{P}} \ .
\eeq

\section{Higher dimensions}
\label{dim}

We briefly discuss how these results generalize to the case of $d>2$ dimensions; we
consider a $d$-dimensional region with $|x_i| \leq a_i/2$ and $i=1,\dots,d$.
For simplicity we restrict our analysis to Dirichlet boundary conditions, since
the cases corresponding to the other boundary conditions can be obtained
in an analogous way.

The Green's function reads
\beq
G^{(D)}(x_1,\dots, x_d; x'_1,\dots,x'_d) &=& \sum_{n_1=1}^\infty \dots \sum_{n_d=1}^\infty 
\frac{1}{\epsilon^{(1D)}_{n_1} + \dots + \epsilon^{(dD)}_{n_d} }  \nonumber \\
&\cdot&\psi^{(1D)}_{n_1}(x_1) \dots \psi^{(d D)}_{n_d}(x_d) 
\psi^{(1D)}_{n_1}(x'_1) \dots \psi^{(d D)}_{n_d}(x'_d) 
\label{greend_dim}
\eeq
where:
\beq
\psi^{(iD)}_{n_i}(x_i) \equiv \sqrt{\frac{2}{a_i}} \ \sin \left( \frac{n_i \pi}{a_i} (x_i+a_i/2)\right) \ \ , \ \ n_i &=& 1,2,\dots  \nonumber 
\eeq
and
\beq
\epsilon^{(iD)}_{n_i} \equiv \frac{n_i^2 \pi^2}{a_i^2} \ \ , \ \ n_i &=& 1,2,\dots  \ .
\eeq

We may now apply Eq.(\ref{eqgd}) on any of the $d$ series contained  in Eq.(\ref{greend_dim});
for instance, if we use it on the $d^{th}$ series we have
\beq
G^{(D)}(x_1,\dots, x_d; x'_1,\dots,x'_d) &=& \sum_{n_1=1}^\infty \dots \sum_{n_{d-1}=1}^\infty 
g_{n_1,\dots , n_{d-1}}^{(D)}(x_d,x'_d) \nonumber \\ 
&\cdot& \psi^{(1D)}_{n_1}(x_1) \dots \psi^{(d-1 \ D)}_{n_{d-1}}(x_{d-1}) 
\psi^{(1D)}_{n_1}(x'_1) \dots \psi^{(d-1 \ D)}_{n_{d-1}}(x'_{d-1})
\label{greend_dim_2}
\eeq
where
\beq
g_{n_1,\dots , n_{d-1}}^{(D)}(x_d,x'_d) &\equiv& \sum_{n_d=1}^\infty \frac{\psi^{(d D)}_{n_d}(x_d) \psi^{(d D)}_{n_d}(x'_d)}{\epsilon^{(dD)}_{n_d}+\Gamma_{n_1,\dots , n_{d-1}}  } \nonumber \\
&=& \frac{ \sinh \left( \sqrt{\Gamma_{n_1,\dots , n_{d-1}}} (x_{d<}+a_d/2)\right) 
\sinh \left(\sqrt{\Gamma_{n_1,\dots , n_{d-1}}} (a_d/2-x_{d>})\right) }{\sqrt{\Gamma_{n_1,\dots , n_{d-1}}}  
\sinh \sqrt{\Gamma_{n_1,\dots , n_{d-1}}} a_d } 
\label{eqgd_dim}
\eeq
and
\beq
\Gamma_{n_1,\dots , n_{d-1}} \equiv \sum_{i=1}^{d-1} \epsilon^{(iD)}_{ni}   \ .
\eeq

The expressions for the sum rules in this case are the analogous of the ones discussed in the previous section, with the appropriate Green's functions and density, which is now 
a function of $d$ variables.

It is interesting to see what happens in the case of a density which depends only
on one direction, for instance $\Sigma = \Sigma(x_d)$. In this case, as for the two 
dimensional case, the eigenfunctions corresponding to the homogeneous directions
can be eliminated from the expressions for the sum rules, using their 
orthogonality. Therefore the sum rules of {\sl any integer order} for
a d-dimensional system of this kind can be expressed in term of $d-1$ infinite series.

\section{Applications}
\label{appl}

We now discuss few applications of the results obtained in the previous sections.

\subsection{Circular annulus}

In Refs.\cite{Amore12,Alvarado11} we have discussed the application of perturbation theory to the 
study of the spectrum of a circular annulus. In particular, in ref.~\cite{Amore12} we have evaluated 
the sum rule of order two for the eigenvalues of the circular annulus of unit external radius and internal radius $r$, using the matrix elements of the conformal density and we have calculated the Casimir energy of the annulus in the limit $r \rightarrow 1^-$.

We will now apply the results of the previous sections to obtain explicit expressions for the sum 
rules for a circular annulus: the function
\beq
f(z) = e^{z+ \frac{1}{2} \log r_{min}}
\label{map}
\eeq
maps the rectangle $\left[ \frac{1}{2} \log r_{min}, -\frac{1}{2} \log r_{min}\right] \times \left[ -\pi, \pi \right]$ onto a circular annulus of external radius $R=1$ and internal radius $r_{min}$. Instead of solving the Helmholtz equation on the annulus, using the conformal map we may solve an equivalent Helmholtz equation for an inhomogeneous medium of density
\beq
\Sigma(x,y) = r_{min} \ e^{2x} 
\eeq
over the rectangle~\cite{Amore12,Alvarado11,Amore10}. Since this density depends only on one coordinate, we will be able to obtain general expressions for the sum rules of the circular annulus in terms of a single series, as we have
pointed out in the previous section.

\begin{figure}
\begin{center}
\bigskip\bigskip\bigskip
\includegraphics[width=6cm]{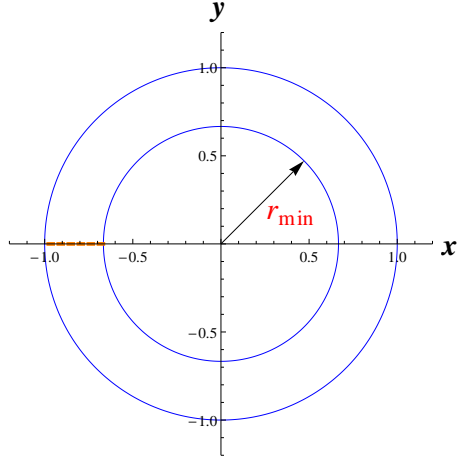}
\caption{Circular annulus obtained acting with the map (\ref{map}) the rectangle $\left[ \frac{1}{2} \log r_{min}, -\frac{1}{2} \log r_{min}\right] \times \left[ -\pi, \pi \right]$. Periodic boundary conditions are imposed on the dashed line.}
\label{Fig_1}
\end{center}
\end{figure}

We will discuss the two cases of an annulus with either Dirichlet or Neumann bc at the border: the first case corresponds to using Dirichlet bc along the $x$ direction and periodic bc along the $y$ direction; the second case corresponds to using Neumann bc along the $x$ direction and periodic bc along the $y$ direction. Notice that using Dirichlet bc along the $y$ direction would correspond to studying an annulus with a transverse cut~\cite{Amore12}.

Applying the general formulas derived in this paper it is possible to obtain explicit expressions
for the sum rules of the circular annulus; for example the sum rule of order two is 
obtained as
\beq
Z^{(DP)}_2 (r_{min})&=& 2 \int_{\log r_{min}}^{-\log r_{min}} dx \ \int_{\log r_{min}}^{x} dx' 
\left[ g_{0,1}^{(DP)}(x,x') \right]^2 \Sigma(x) \Sigma(x') \nonumber \\ 
&+& 2 \sum_{n_y=1}^\infty \sum_{u_y=1}^2 \int_{\log r_{min}}^{-\log r_{min}} dx \ \int_{\log r_{min}}^{x} dx' \left[ g_{n_y,u_y}^{(DP)}(x,x') \right]^2 \Sigma(x) \Sigma(x') \nonumber
\eeq

Using the explicit expression for $ g_{0,1}^{(DP)}(x,x')$ and performing the integrations
one obtains
\beq
Z^{(DP)}_2 (r_{min}) &=& \frac{r_{min}^4 \log^4(r_{min})}{4-4 r_{min}^4}-\frac{5}{64} 
\left(r_{min}^4-1\right) \log^2(r_{min})+\frac{1}{16} \left(r_{min}^2-1\right)^2  
\log(r_{min}) \nonumber \\
&+& \frac{r_{min}^4 \log^5(r_{min})}{2 \left(r_{min}^2-1\right)^2} 
- \frac{\log^3(r_{min})}{144 \left(r_{min}^2+1\right)^2} \left(26 r^8+73 r_{min}^6+62 r_{min}^4 \right. \nonumber \\
&+& \left. 73 r_{min}^2-3 \pi ^2 \left(r_{min}^2+1\right)^2 \left(r_{min}^4+1\right)+26\right) \nonumber \\
&+& \sum_{n=3}^\infty \left[ \frac{n \left(n^2+5\right) \left(r_{min}^4-1\right) 
r_{min}^{4 n}}{8\left(n^2-4\right) \left(n^2-1\right)^2 \left(r_{min}^{2n}-1\right)^2}+\frac{n^2 \left(r_{min}^2-1\right)^2 r_{min}^{2 n}}{2 \left(n^2-1\right)^2 \left(r_{min}^{2 n}-1\right)^2} \right]
\label{zeta2dpannulusa}
\eeq

This expression may be cast in terms of a more rapidly convergent series
observing that $0<r_{min}<1$ and expanding the factor $1/\left(r_{min}^{2 n}-1\right)^2$ 
in powers of $r$. After performing the summation over $n$ one is left
with the new series:
\beq
Z^{(DP)}_2 (r_{min}) &=& -\frac{\left(r_{min}^4-1\right)^2 \left(r_{min}^4+1\right) 
{\rm Li}_2\left(r_{min}^4\right)}{16 r_{min}^4}
+\frac{\left(r_{min}^2-1\right)^2 \left(r_{min}^4+1\right) {\rm Li}_2\left(r_{min}^2\right)}{8 r_{min}^2} \nonumber \\
&+& \frac{r^4\log (r_{min})}{4 \left(1-r_{min}^4\right)} - 
\frac{5 \left(r_{min}^4-1\right)}{64 \log(r_{min})}
+\frac{\left(r_{min}^2-1\right)^2}{16 \log^2(r_{min})} \nonumber \\
&-& \frac{\left(r_{min}^2-1\right)^3 \left(r_{min}^2+1\right) \log\left(1-r_{min}^2\right)}{8 r_{min}^2} \nonumber \\
&-& \frac{\left(r_{min}^4-1\right)^3 \left(r_{min}^8+r_{min}^4+1\right) \log \left(1-r_{min}^4\right)}{16 r_{min}^8} + \frac{r_{min}^4 \log^2(r_{min})}{2 \left(r_{min}^2-1\right)^2} \nonumber \\
&+& \frac{1}{576} \left[71 r_{min}^{12}-202 r_{min}^8+274 r_{min}^6-297 r_{min}^4+12 \pi^2
   \left(r_{min}^4+1\right) \right. \nonumber \\
&+& \left. \frac{36}{r_{min}^4}-66 r_{min}^2+\frac{4 \left(5 r_{min}^4-54
   r_{min}^2-27\right)}{\left(r_{min}^2+1\right)^2}\right] \nonumber \\
&+& \sum_{j=1}^\infty (j+1) \left[
\frac{1}{16} \left(r_{min}^4-1\right) \left(r_{min}^{4 j}-1\right) r_{min}^{-2 j} 
{\rm Li}_2\left(r_{min}^{2j}\right) \right. \nonumber \\
&-& \left. \frac{1}{16} \left(r_{min}^4-1\right) \left(r_{min}^{4 j+8}-1\right) r_{min}^{-2 (j+2)} 
{\rm Li}_2\left(r_{min}^{2j+4}\right) \right. \nonumber \\
&+& \left. \frac{1}{8} \left(r_{min}^2-1\right)^2 \left(r_{min}^{4j+4}+1\right) r_{min}^{-2 (j+1)} 
{\rm Li}_2\left(r_{min}^{2j+2}\right)  \right. \nonumber \\
&+& \left. \frac{1}{16} \left(r_{min}^4-1\right) \left(r_{min}^{2 j}-1\right)^2 \left(r_{min}^{2 j}+r_{min}^{4 j}+1\right) r_{min}^{-4 j} \log \left(1-r_{min}^{2j}\right) \right. \nonumber \\
&-& \left. \frac{1}{16} \left(r_{min}^4-1\right) \left(r_{min}^{2 j+4}-1\right)^2 
\left(r_{min}^{2 j+4}+r_{min}^{4 j+8}+1\right) r_{min}^{-4 (j+2)} \log \left(1-r_{min}^{2j+4}\right) \right. \nonumber \\
&-& \left. \frac{1}{8} \left(r_{min}^2-1\right)^2 \left(r_{min}^{4j+4}-1\right) r_{min}^{-2 (j+1)} 
\log \left(1-r_{min}^{2j+2}\right) \right. \nonumber \\
&+& \left. \frac{r^{-2(j+2)}}{576} \left(-3 \left(r_{min}^2-1\right)^2 \left(r_{min}^4-4
   r_{min}^2+1\right) r_{min}^{4 j+4} \right. \right. \nonumber \\
&+& \left. \left. \left(r_{min}^2-1\right)^2 \left(71 r_{min}^8+142 r_{min}^6+14
   r_{min}^4+142 r_{min}^2+71\right) r_{min}^{6 j+4} \right.\right. \nonumber \\
&+& \left.\left. 36 \left(r_{min}^4-1\right)^2\right) \right]
\label{zeta2dpannulusb}
\eeq

Clearly the rate of convergence of this series increases for $r_{min} \rightarrow 0^+$:
in this limit one obtains the behavior
\beq
Z^{(DP)}_2 (r_{min}) &\approx& \left(\frac{\pi ^2}{48}-\frac{5}{32}\right) +
\left(\frac{1}{16 \log^2(r_{min})}+\frac{5}{64 \log (r_{min})}\right) \nonumber \\
&-& r_{min}^2 \left(\frac{1}{8 \log^2(r_{min})}+\frac{7}{48}\right)+ \dots
\eeq

Notice that 
\beq
\lim_{r_{min}\rightarrow 0^+} Z^{(DP)}_2 (r_{min}) = \left(\frac{\pi ^2}{48}-\frac{5}{32}\right)
\approx 0.04936675836
\eeq
is the Dirichlet sum rule of order two for the unit circle.

Notice also that the derivative of $Z^{(DP)}_2 (r_{min})$ diverges at $r=0$:
\beq
\lim_{r_{min}\rightarrow 0^+} \frac{d Z^{(DP)}_2 (r_{min})}{dr_{min}} = -\frac{1}{8 r_{min} \log ^3(r_{min})}-\frac{5}{64 r_{min} \log ^2(r_{min})} =
- \infty
\eeq

In a similar way we may obtain the sum rule of higher orders: although we have proved that
these sum rules will involve a single series, since the density depends only on one direction,
their expressions become  lengthier and we do not see any advantage in reporting them here.

Instead, we investigate explicitly the limit of an infinitesimal hole, as done in the previous case:
\beq
Z^{(DP)}_3 (r_{min}) &\approx& \left(\frac{\zeta (3)}{32}+\frac{35}{768}-\frac{\pi ^2}{128}\right) \nonumber \\
&+& \left(\frac{1}{64 \log ^3(r_{min})}+\frac{15}{512 \log^2(r_{min})}+\frac{23}{1152 \log(r_{min})} \right) \nonumber \\
&-& r_{min}^2 \left(\frac{3}{64 \log^3(r_{min})} +
\frac{15}{512 \log^2(r_{min})}+\frac{19}{1536}\right) + \dots
\eeq

Notice that
\beq
\lim_{r_{min}\rightarrow 0^+} Z^{(DP)}_3 (r_{min}) = \frac{\zeta (3)}{32}+\frac{35}{768}-\frac{\pi ^2}{128} \approx
0.006030910507
\eeq
is the corresponding Dirichlet sum rule for a unit circle. Once again we see that the derivative
of $ Z^{(DP)}_3 (r_{min})$ diverges at $r_{min}=0$.

The last case that we examine is sum rule of order four: for $r_{min} \rightarrow 0^+$ it behaves as
\beq
Z^{(DP)}_4 (r_{min}) &\approx& \left( -\frac{\zeta (3)}{64}-\frac{3491}{110592}
+\frac{5 \pi^2}{1152}+\frac{\pi ^4}{11520} \right) \nonumber \\
&+& \left( \frac{1}{256 \log ^4(r_{min})}+\frac{5}{512 \log ^3(r_{min})}
+\frac{2147}{221184 \log ^2(r_{min})}+\frac{677}{147456 \log (r_{min})} \right) \nonumber \\
&-& r_{min}^2 \left( \frac{1}{64 \log ^4(r_{min})}+\frac{5}{256 \log ^3(r_{min})}+\frac{23}{3456 \log^2(r_{min})} +\frac{149}{138240}\right) + \dots
\eeq

We have
\beq
\lim_{r_{min}\rightarrow 0^+} Z^{(DP)}_4 (r_{min}) &=& \left( -\frac{\zeta (3)}{64}-\frac{3491}{110592}
+\frac{5 \pi^2}{1152}+\frac{\pi ^4}{11520} \right) \nonumber \\
&\approx& 0.0009438572210 \ ,
\eeq
which is the corresponding sum rule for a unit circle with Dirichlet bc.
The derivative of $ Z^{(DP)}_4 (r_{min})$ diverges at $r=0$.

We now discuss the case of Neumann boundary conditions: the evaluation of these sum rules
requires using the same expressions used for the case od Dirichlet bc, with 
$g_{n_y,u_y}^{(NP)}(x,x')$ instead of $g_{n_y,u_y}^{(DP)}(x,x')$.

In the case of Neumann bc and of a circular annulus with an infinitesimal hole we find
\beq
Z^{(NP)}_2 (r_{min}) &\approx& \left( \frac{\log ^2(r_{min})}{36}+\frac{\log(r_{min})}{8} \right) +
\left( \frac{5 \pi ^2}{48}-\frac{49}{96} \right)
\nonumber \\
&+& \left( \frac{7}{32 \log ^2(r_{min})} +\frac{25}{64 \log (r_{min})} \right) 
   \nonumber \\  
&+& r_{min}^2 \left( \frac{77}{48}-\frac{1}{72} \log^2(r_{min})-\frac{7}{16 \log^2(r_{min})} \right)  + \dots
\eeq

The logarithmic divergence of this expression signals the presence of eigenvalues of
infinitesimal magnitude as $r_{min} \rightarrow 0^+$. 

It interesting to study the sum rules of a circular annulus with an infinitesimal 
hole and mixed boundary conditions, either Neumann-Dirichlet or Dirichlet-Neumann 
on the internal and external borders respectively.

For the Neumann-Dirichlet case we find the sum rule of order two:
\beq
Z^{(NDP)}_2 (r_{min}) &\approx& \left(\frac{\pi^2}{48}-\frac{5}{32}\right) -\frac{5 r_{min}^2}{48} 
+ r_{min}^4 \left( \frac{5 \pi^2}{48}-\frac{143}{288} \right) \nonumber \\
&+& r_{min}^4 \log(r_{min}) \left( \frac{3}{4} \log(r_{min})+\frac{11}{8} \right) \nonumber \\ 
&-& \frac{6377 r_{min}^6}{2880} - r_{min}^6 \log (r_{min}) \left(\log(r_{min})+4 \right) + \dots 
\eeq

We notice that $Z^{(NDP)}_2 (r_{min})$ tends to the corresponding sum rule for the unit circle
with Dirichlet boundary conditions and that  the first three derivatives of $Z^{(NDP)}_2 (r_{min})$ are finite at $r_{min}=0^+$ and in particular that $\lim_{r_{min}\rightarrow 0^+} dZ^{(NDP)}_2 (r_{min})/dr_{min} =0$: however the presence of a term
$r_{min}^4 \log(r_{min})$ implies that $\lim_{r_{min}\rightarrow 0^+} d^4Z^{(NDP)}_2(r_{min})/dr_{min}^4 = \infty$.

Similarly we find the sum rule of order three:
\beq
Z^{(NDP)}_3 (r_{min}) &\approx& \left(\frac{\zeta (3)}{32}-\frac{\pi ^2}{128}+\frac{35}{768}\right) -\frac{71 r_{min}^2}{1536} + \dots \\
&-& \frac{1781 r_{min}^4}{23040}-\frac{19}{64} r_{min}^4 \log (r_{min}) + 
r_{min}^6 \left( -\frac{7  \zeta (3)}{32}-\frac{19 \pi ^2 }{128}
+\frac{162319 }{92160} \right) \nonumber \\
&-& r_{min}^6 \log(r_{min}) \left( \frac{1}{8} \log^2(r_{min}) + \frac{57}{32} \log(r_{min}) + \frac{85}{64} \right) + \dots
\eeq   
  
In this case we observe that $Z^{(NDP)}_3 (r_{min})$ tends to the corresponding sum rule for the unit circle
with Dirichlet boundary conditions and that  the first three derivatives of $Z^{(NDP)}_3 (r_{min})$ are finite at $r_{min}=0^+$ and in particular that $\lim_{r_{min}\rightarrow 0^+} dZ^{(NDP)}_3 (r_{min})/dr_{min} =0$: however the presence of a term
$r_{min}^4 \log(r_{min})$ implies that $\lim_{r_{min}\rightarrow 0^+} d^4Z^{(NDP)}_3 (r_{min})/dr_{min}^4 = \infty$.

For the sum rule of order four we have   
\beq
Z^{(NDP)}_4 (r_{min}) &\approx& \left( -\frac{\zeta (3)}{64} + \frac{\pi ^4}{11520} + \frac{5 \pi^2}{1152} - 
\frac{3491}{110592} \right) -\frac{1691 r_{min}^2}{138240} \\
&+& \frac{40489 r_{min}^4}{829440}-\frac{109 r_{min}^4 \log (r_{min})}{2304} \nonumber \\
&+& \frac{13140797 r_{min}^6}{116121600}-\frac{5}{96} r_{min}^6 \log ^2(r_{min})+\frac{571 r_{min}^6 \log(r_{min})}{1152}
\dots
\eeq

Once again we observe that $Z^{(NDP)}_4 (r_{min})$ tends to the corresponding sum rule for the unit circle
with Dirichlet boundary conditions and that  the first three derivatives of $Z^{(NDP)}_4 (r_{min})$ are finite at $r_{min}=0^+$ and in particular that $\lim_{r_{min}\rightarrow 0^+} dZ^{(NDP)}_4 (r_{min})/dr_{min} =0$: however the presence of a term
$r_{min}^4 \log(r_{min})$ implies that $\lim_{r_{min}\rightarrow 0^+} d^4Z^{(NDP)}(4)/dr_{min}^4 = \infty$.

A different behavior is observed for the sum rules of the Dirichlet-Neumann case: for example, for the
sum rule of order two we have
\beq
Z^{(DNP)}_2 (r_{min}) &\approx& \left( \frac{\log ^2(r_{min})}{4}+\frac{3 \log (r_{min})}{8} \right)+
\left( \frac{5 \pi^2}{48}-\frac{19}{32}\right) -\frac{85 r_{min}^2}{48} + \dots 
\eeq
which diverges for $r_{min} \rightarrow 0^+$, as in the analogous case of Neumann-Neumann boundary conditions.

\subsection{Circular sector}

\begin{figure}
\begin{center}
\bigskip\bigskip\bigskip
\includegraphics[width=6cm]{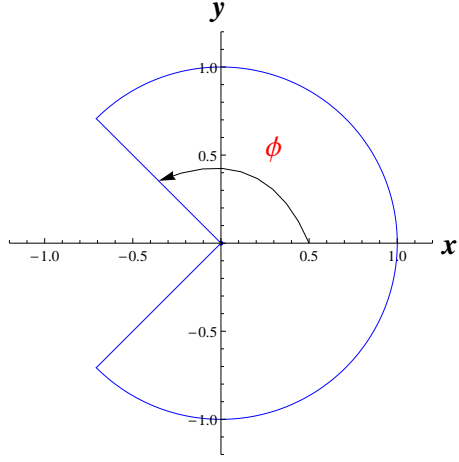}
\caption{Circular sector obtained acting with the map (\ref{map}) the rectangle 
$\left[ \frac{1}{2} \log r, -\frac{1}{2} \log r\right] \times \left[ -\phi, \phi \right]$, with
$r \rightarrow 0^+$.}
\label{Fig_2}
\end{center}
\end{figure}

We consider the circular sector represented in Fig.\ref{Fig_2}: this domain is obtained
applying the map (\ref{map}) to the rectangle $\left[ \frac{1}{2} \log r, -\frac{1}{2} \log r\right] 
\times \left[ -\phi, \phi \right]$, where $r \rightarrow 0^+$. Therefore it is natural to extend
the analysis that we have done for the case of the circular annulus to this case.

The eigenfunctions of the negative laplacian on this domain  are
\beq
\Psi_{n,k}(r,\theta) = N_{nk} J_{\frac{n\pi}{2\phi}}(\alpha_{nk} r) \sin \left[\frac{n\pi}{2\phi} (\phi+\theta)\right]
\eeq
where $N_{nk}$ is a normalization constant:
\beq
N_{nk} = \frac{\sqrt{2}}{\sqrt{\phi  \left(J_{\frac{n \pi }{2 \phi }}(\alpha_{nk} ){}^2-J_{\frac{n \pi}{2 \phi }-1}(\alpha_{nk}) J_{\frac{\pi  n}{2 \phi }+1}(\alpha_{nk} )\right)}} \ .
\eeq

The eigenvalues of the negative domain are then:
\beq
E_{nk} = \alpha_{nk}^2 \ ,
\eeq
where $\alpha_{nk}$ is the $k^{th}$ zero of $J_{\frac{n\pi}{2\phi}}(x)$.

We first discuss the case of Dirichlet boundary conditions at the borders of the circular sector; the general 
rules that we have derived in Section \ref{exact} can be applied straightforwardly.

The sum rule of order two is
\beq
Z_2^{(D)}(\phi) &=& \frac{1}{2} \sum_{n=1}^\infty \frac{1}{\left(\frac{\pi  n}{\phi }+2\right)^2 
\left(\frac{\pi  n}{\phi}+4\right)} \nonumber \\
&=& \frac{\phi ^2}{4 \pi ^2} \ \psi ^{(1)}\left(\frac{2 \phi }{\pi }+1\right) 
+ \frac{\phi}{8 \pi } \ \psi^{(0)}\left(\frac{2 \phi }{\pi }+1\right)
- \frac{\phi}{8 \pi } \  \psi^{(0)}\left(\frac{4 \phi }{\pi }+1\right)
\eeq
where $\psi ^{(n)}\left(z\right)\equiv (-1)^{n+1} n! \sum_{k=0}^\infty \frac{1}{(k+z)^{n+1}}$ is a
polygamma function.

The sum rule of order three is
\beq
Z_3^{(D)}(\phi) &=& \sum_{n=1}^\infty \frac{1}{\left(\frac{\pi  n}{\phi }+2\right)^3 
\left(\frac{\pi  n}{\phi }+4\right) \left(\frac{\pi  n}{\phi }+6\right)} \nonumber \\
&=& - \frac{\phi^3}{16 \pi ^3} \ \psi ^{(2)}\left(\frac{2 \phi}{\pi }+1\right)
- \frac{3 \phi^2 }{32 \pi ^2} \ \psi ^{(1)}\left(\frac{2 \phi }{\pi }+1\right)
- \frac{7 \phi}{128 \pi } \ \psi^{(0)}\left(\frac{2 \phi }{\pi }+1\right) \nonumber \\
&+& \frac{\phi}{16 \pi } \ \psi^{(0)}\left(\frac{4 \phi }{\pi }+1\right)
-\frac{\phi}{128 \pi } \ \psi^{(0)}\left(\frac{6 \phi }{\pi }+1\right)
\eeq

The sum rule of order four is
\beq
Z_4^{(D)}(\phi) &=& \frac{1}{2} \sum_{n=1}^\infty \frac{\frac{5 \pi  n}{\phi }+22}{\left(\frac{\pi  n}{\phi }+2\right)^4 \left(\frac{\pi n}{\phi }+4\right)^2 \left(\frac{\pi  n}{\phi }+6\right) \left(\frac{\pi  n}{\phi}+8\right)} \nonumber \\
&=& \frac{\phi^4}{96 \pi ^4} \ \psi ^{(3)}\left(\frac{2 \phi }{\pi }+1\right)
+\frac{\phi^3}{32 \pi ^3} \ \psi ^{(2)}\left(\frac{2 \phi }{\pi }+1\right)
+\frac{17 \phi^2}{384 \pi ^2} \ \psi^{(1)}\left(\frac{2 \phi }{\pi }+1\right) \nonumber \\
&+& \frac{\phi^2}{128 \pi ^2} \ \psi^{(1)}\left(\frac{4 \phi }{\pi }+1\right) 
+\frac{127 \phi}{4608 \pi } \ \psi^{(0)}\left(\frac{2 \phi }{\pi }+1\right)
-\frac{15 \phi}{512 \pi } \ \psi^{(0)}\left(\frac{4 \phi }{\pi }+1\right) \nonumber \\
&+& \frac{\phi}{512 \pi } \ \psi^{(0)}\left(\frac{6 \phi }{\pi }+1\right)
-\frac{\phi}{4608 \pi } \ \psi^{(0)}\left(\frac{8 \phi }{\pi }+1\right)
\eeq

\begin{table}[tbp]
\caption{Dirichlet sum rules for circular sector at specific angular values}
\bigskip
\label{table1}
\begin{center}
\begin{tabular}{|c|c|}
\hline
$\phi$ & $Z_2^{(D)}(\phi)$ \\
\hline
$\frac{\pi }{4}$   & $-\frac{1}{32}+\frac{\pi ^2}{128}-\frac{\log (4)}{32}$ \\
$\frac{\pi }{2}$   & $\frac{\pi ^2}{96}-\frac{3}{32}$ \\
$\frac{3 \pi }{4}$ & $-\frac{35}{64}+\frac{9 \pi ^2}{128}-\frac{3 \log (4)}{32}$ \\
$\pi$              & $\frac{\pi^2}{24}-\frac{37}{96}$  \\
\hline
$\phi$ & $Z_3^{(D)}(\phi)$ \\
\hline
$\frac{\pi }{4}$   & $\frac{7 \zeta (3)}{512}-\frac{7}{768}-\frac{3 \pi ^2}{1024}+\frac{\log (4)}{64}$ \\
$\frac{\pi }{2}$   & $\frac{\zeta (3)}{64}+\frac{31}{1536}-\frac{\pi ^2}{256}$ \\
$\frac{3 \pi }{4}$ & $\frac{189 \zeta (3)}{512}-\frac{6653}{26880}-\frac{27 \pi^2}{1024}+\frac{3 \log (4)}{64}$ \\
$\pi$              & $\frac{\zeta (3)}{8}+\frac{43}{7680}-\frac{\pi ^2}{64}$ \\
\hline
$\phi$ & $Z_4^{(D)}(\phi)$ \\
\hline
$\frac{\pi }{4}$   & $-\frac{7 \zeta (3)}{1024}+\frac{1}{36864}+\frac{3 \pi ^2}{2048}+\frac{\pi^4}{24576}
-\frac{17 \log (4)}{2304}$ \\
$\frac{\pi }{2}$   & $-\frac{\zeta (3)}{128}-\frac{1795}{110592}+\frac{5 \pi ^2}{2304}+\frac{\pi^4}{23040}$ \\
$\frac{3 \pi }{4}$ & $-\frac{189 \zeta (3)}{1024}-\frac{256171}{1290240}+\frac{27 \pi^2}{2048}+\frac{27 \pi ^4}{8192}-\frac{17 \log (4)}{768}$ \\
$\pi$              & $-\frac{\zeta (3)}{16}-\frac{33569}{430080}+\frac{5 \pi ^2}{576}+\frac{\pi^4}{1440}$ \\
\hline
\end{tabular}
\end{center}
\bigskip\bigskip
\end{table}

\subsection{Circular annulus with inhomogenous density}

We now discuss the case of a circular annulus with a density which depends only on the radial 
coordinate. For simplicity we assume an annulus of internal radius $r_{min}$ and external radius $r_{max} =1$
and with density 
\beq
\rho(r) = \frac{(b+2) \left(r_{min}^2-1\right) r^b}{2 \left(r_{min}^{b+2}-1\right)} \ .
\eeq
These drums all have the same mass $M= 2\pi \int_{r_{min}}^1 \rho(r) r dr = \pi (1-r_{min}^2)$.

The eigenmodes of these annular membranes are the eigensolutions of the Helmholtz equation
\beq
\left(-\Delta \right) \psi_n (r,\theta) = E_n \rho(r) \psi_n(r,\theta)
\label{radialannulus1}
\eeq 
where $r \in (r_{min},1)$ and $\theta \in (0,2\pi)$. 

Using the conformal map (\ref{map}) we convert the Helmholtz equation (\ref{radialannulus1}) into
\beq
-\left(\frac{\partial}{\partial x^2} +\frac{\partial}{\partial y^2} \right) \Phi_n (x,y) = E_n 
\tilde{\Sigma}(x)  \Phi_n(x,y)
\label{radialannulus2}
\eeq 
where $x \in (\log(r_{min})/2,-\log(r_{min})/2)$ and $y \in (-\pi,\pi)$ and 
\beq
\tilde{\Sigma}(x) \equiv  (r_{min} e^{2x}) \rho(\sqrt{r_{min}} e^x) 
\eeq

It is easy to check that 
\beq
\left. \tilde{\Sigma}(x)  \right|_{b=-2+\delta} = \left. \tilde{\Sigma}(-x)  \right|_{b=-2-\delta}  \nonumber
\eeq
and therefore we conclude that the annuli with exponents $b= -2 \pm \delta$ are {\sl isospectral}. Using $\delta=2$
we see that the uniform annulus is isospectral to the annulus with density $\rho(r) = r_{min}^2/r^4$ (this
case has been studied by Gottlieb in ref.~\cite{Gottlieb06b})\footnote{A more general class of radially isospectral annular membrane is discussed  in ref.~\cite{Gottlieb06a}.}.

The sum rule of order two for this inhomogeneous annulus is
\beq
Z_2^{(D)}(b) &=& \frac{\left(r_{min}^2-1\right)^2}{8 (b+2)^4
   \left(r_{min}^{b+2}-1\right)^2 \log ^2(r_{min})} \nonumber \\
&\cdot&    \left[8 \left(r_{min}^{b+2}-1\right)^2+(b+2) \log (r_{min}) \left(-5 r_{min}^{2
   b+4}+(b+2) \left(r_{min}^{2 b+4}+1\right) \log (r_{min})+5\right)\right] \nonumber \\
   &+& \sum_{n=1}^\infty \frac{\mathcal{N}_n}{\mathcal{D}_n}
\eeq
where we have defined
\beq
\mathcal{N}_n &\equiv& -\left(r_{min}^2-1\right)^2 \left(-32 n^2 \left(n^2-(b+2)^2\right) r_{min}^{b+2}
+\left((b+2)^2-4 n^2\right)^2 r_{min}^{2 b+4}+\left((b+2)^2-4 n^2\right)^2\right) \nonumber \\
&+& \frac{1}{2} (b+2) \left(r_{min}^2-1\right)^2 r_{min}^{-2 n} \left((b+n+2) (b+2 n+2)^2 r_{min}^{2 b+4}+(b-2 n+2)^2
   (b-n+2)\right) \nonumber \\
&+& \frac{1}{2} (b+2) \left(r_{min}^2-1\right)^2 r_{min}^{2 n} \left((b-2 n+2)^2 (b-n+2) r_{min}^{2 b+4}+(b+n+2) (b+2 n+2)^2\right) \\
\mathcal{D}_n &\equiv& 2 \left(b^2+4 b-4 n^2+4\right)^2 \left(b^2+4 b-n^2+4\right) \left(r_{min}^{b+2}-1\right)^2
   \left(r_{min}^{-n}-r_{min}^n\right)^2
\eeq

In particular for $b=-2$ and $r_{min} \rightarrow 0^+$ we find:
\beq
Z_2^{(D)}(-2) &\approx& \frac{\log^2(r_{min})}{360}-\frac{\zeta (3)}{8 \log (r_{min})}-\frac{\pi ^4}{360 \log^2(r_{min})} + \dots
\label{asym}
\eeq

\begin{figure}
\begin{center}
\bigskip\bigskip\bigskip
\includegraphics[width=8cm]{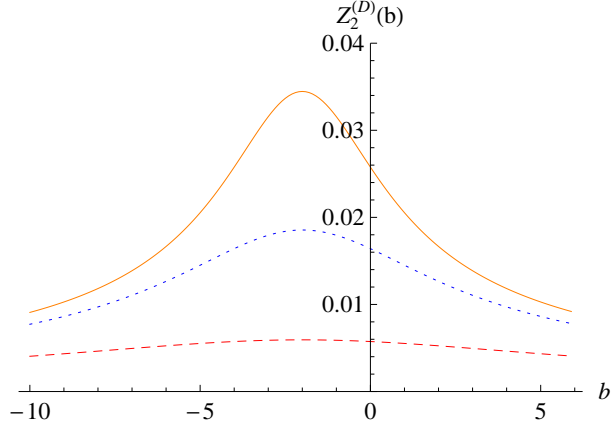}
\caption{Sum rule of order $2$ for a circular annulus with density $\rho(r) = \frac{(b+2) \left(r_{min}^2-1\right)
   r^b}{2\left(r_{min}^{b+2}-1\right)}$.
The curves from top to bottom correspond to $r_{min}=0.1, 0.25, 0.5$.
}
\label{Fig_3}
\end{center}
\end{figure}

In Fig.\ref{Fig_3} we plot the sum rule $Z_2^{(D)}(b)$ at three different values of $r_{min}$ as a functions of $b$.
The curve is symmetric with respect to the axis $b=-2$, as a result of the isospectrality discussed earlier.

\begin{figure}
\begin{center}
\bigskip\bigskip\bigskip
\includegraphics[width=8cm]{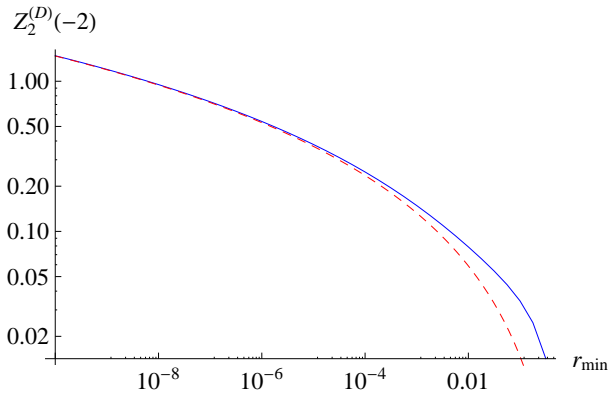}
\caption{$Z_2^{(D)}(-2)$ as a function of $r_{min}$ for a circular annulus with density $\rho(r) =\frac{r_{min}^2-1}{2\log(r_{min})} \frac{1}{r^2}$. The solid line is the exact result, whereas the dashed line is the
asymptotic formula (\ref{asym}).}
\label{Fig_4}
\end{center}
\end{figure}

In Fig.\ref{Fig_4} we plot the sum rule $Z_2^{(D)}(-2)$ as a function of $r_{min}$ (solid line); the
dashed line is the asymptotic behavior of eq.(\ref{asym}).

\section{Conclusions}
\label{concl}

We have extended the results of Ref.~\cite{Amore13a} to two dimensional domains of arbitrary shape and
density; we have derived a general expression for the sum rule of integer order $n$ in terms of the
Green's functions of the homogeneous problem and we have proved that the sum rule of any integer order 
can be expressed in terms of a single series for the case in which the inhomogeneity is only along a direction. We have also discussed the generalization of these results to higher dimensions.

As an application of the formulas that we have derived, we have calculated the sum rules for an homogeneous
circular annulus, for a homogeneous circular sector and for a circular annulus with a radially varying density.
We have found explicit expressions for the sum rules of a circular annulus subject to different boundary conditions and with an infinitesimal internal hole: in the case of Dirichlet boundary conditions on the
external border we have proved that the sum rules tend to the corresponding sum rules for the unit circle.

The possibility of obtaining exact expressions for the sum rules of the eigenvalues of the negative
laplacian on arbitrary domains may be exploited to estimate the higher order coefficients
in Weyl's asymptotic law. This aspect would provide a natural extension of the present work and we
plan to consider it in the future. As we have discussed in \cite{Amore13a}, following \cite{Berry86}, 
the sum rules may be used to provide rigorous upper and lower bounds to the lowest eigenvalue.

\section*{Acknowledgements}
This research was supported by the Sistema Nacional de Investigadores (M\'exico).

\end{document}